\font\ninerm=cmr9  \font\sixrm=cmr6
\font\ninei=cmmi9  \font\sixi=cmmi6
\font\ninesy=cmsy9  \font\sixsy=cmsy6

\font\ninebf=cmbx9  \font\sixbf=cmbx6

\font\twelverm=cmr12 \font\twelvei=cmmi12 \font\twelvesy=cmsy12
\font\twelveit=cmti12 \font\twelvesl=cmsl12 \font\twelvebf=cmbx12
\font\twelvett=cmtt12 

\font\bgp=cmbx12 scaled\magstep1

\def\twelvepoint{\def\rm{\fam0\twelverm}%
  \textfont0=\twelverm \scriptfont0=\ninerm \scriptscriptfont0=\sixrm
  \textfont1=\twelvei \scriptfont1=\ninei \scriptscriptfont1=\sixi
  \textfont2=\twelvesy \scriptfont2=\ninesy \scriptscriptfont2=\sixsy
  \textfont3=\tenex \scriptfont3=\tenex \scriptscriptfont3=\tenex
  \def\it{\fam\itfam\twelveit}%
  \textfont\itfam=\twelveit
  \def\sl{\fam\slfam\twelvesl}%
  \textfont\slfam=\twelvesl
  \def\bf{\fam\bffam\twelvebf}%
  \textfont\bffam=\twelvebf \scriptfont\bffam=\ninebf
   \scriptscriptfont\bffam=\sixbf
  \def\tt{\fam\ttfam\twelvett}%
  \textfont\ttfam=\twelvett
}


%
\twelvepoint\rm
\hsize= 6.5truein
\vsize= 8.50truein
\hoffset= 0.0truein
\voffset= 0.0truein
\lineskip= 2pt
\lineskiplimit= 2pt
\overfullrule=0pt
\tolerance=2000
\topskip= 0pt
\baselineskip=15pt
\parindent=0.4truein
\parskip=0pt plus1pt
\footline={\hss\twelverm\folio\hss}
\def\medskip{\vskip6pt plus2pt minus2pt}
\def\bigskip{\vskip12pt plus4pt minus4pt}
\def\smallskip{\vskip3pt plus1pt minus1pt}
\def\ep{\epsilon}
\def\up{\uparrow}
\def\down{\downarrow}

\def\rt2{{1\over\sqrt2}}
\def\o2{{1\over2}}
\def\exp{\mathop{\rm e}\nolimits}
\def\cos{\mathop{\rm cos}\nolimits}
\def\sin{\mathop{\rm sin}\nolimits}
\centerline{\bgp Inter-Band Pairing Theory of Superconductivity}
\bigskip
\centerline{Jamil Tahir-Kheli}
\centerline{\it First Principles Research, Inc.}
\centerline{\it 8391 Beverly Blvd., Suite \#171, Los Angeles, CA 90048}
\bigskip
\noindent\centerline{\bf ABSTRACT}\medskip
A model for high temperature superconductors based on the idea of
Cooper pairs comprised of electrons from {\it different}
bands is
studied. We propose that the two bands relevant for the cuprates are
comprised of Cu $d_{x^2-y^2}$, $d_{z^2}$, planar O $p_\sigma$, and
apical O $p_z$ orbitals. Along the diagonal, $k_x=k_y$ in the
Brillouin zone, the two band Fermi surfaces may cross. We associate
the optimal doping for the highest T$_c$ with this point because only
in the vicinity of this touching point are inter-band Cooper pairs
energetically possible. Due to the lack of time reversal
invariance of an inter-band Cooper pair with itself, the standard
interpretation of Josephson tunneling is altered such that the
detailed nature of the single particle tunneling matrix elements
contributes to the supercurrent. The $d_{x^2-y^2}$ gap observations
from Josephson tunneling are shown to arise from our model with
pairing due to phonons. A Hubbard model is written down for the two
bands at the Fermi energy with realistic parameters for
La$_{1.85}$Sr$_{0.15}$CuO$_4$. 
The anomalous normal state
features in the nmr are calculated and qualitatively explained as due
to the character of the two bands in the vicinity of the crossing
point. The Hall effect is calculated using standard Bloch-Boltzmann
transport theory. The observed strong temperature dependence of the
Hall coefficient is reproduced and is due to the strong reshaping of
the current carrying band Fermi surface
due to band repulsion with the other band
for dopings very close to the Fermi surface touching
point. Reasonable quantitative agreement is also obtained for the nmr
and Hall effect.
A linear resistivity at optimal
doping is expected due to the proximity of the second band 
in k space which can
strongly relax the current and the ``smallness'' of the current
carrying Fermi surface.
\bigskip
\noindent \hbox to \hsize{
\hfil
{\tt cond-mat/9711170}}
\bigskip
\noindent {\it Submitted to Phys. Rev. B}
\footnote{}{\ninerm A PostScript version of this paper is available 
for download
at http://www.firstprinciples.com.}
\vfil\eject
%
%
\centerline{\bf I. INTRODUCTION}

In a previous publication the author proposed an Inter-Band Pairing (IBP)
model for high temperature superconductors.$^1$
We suggested that the
fundamental idea of Cooper pairs composed of electrons in 
states $k\up$
and $-k\down$ be retained but rather than $k\up$ and $-k\down$ being
from the same band, we consider pairs such that $k\up$ and $-k\down$
come from {\it different} 
bands at the Fermi surface. Such Cooper pairs, if
they exist, are not time reversal invariant with themselves in
contrast to all BCS-like (Intra-Band) Cooper pairs. The full
Hamiltonian is time reversal invariant, though. We showed that with
such Inter-Band pairs, the orbital character of the two bands plays a
significant role in all Josephson tunneling experiments. The phase
difference across a junction is no longer simply the phase difference
of the two gap functions but also includes a contribution due to the
phases of the hopping matrix elements across the junction. With
BCS-like Cooper pairs, the product of the 
single particle tunneling matrix
elements is mod-squared due to the time reversal invariance of the
pair. This is not the case with IBP. We suggest that the observed$^{2,3}$
$d_{x^2-y^2}$
character of the Josephson tunneling
can be resolved by an ``s-like gap'' coupled
with a ``$d_{x^2-y^2}$'' Cooper pair. These issues will be explained in detail
below.

\hbox{\indent Clearly, in order to 
have any possibility of creating an Inter-Band Cooper pair,\hfil} 
\noindent $(k\up_U,-k\down_L)$ where U, L are labels for the two
distinct bands (eventually taken to represent the Upper and Lower bands), 
the single particle energies $\ep^{(U)}_k=\ep^{(U)}_{-k}$,
$\ep^{(L)}_k=\ep^{(L)}_{-k}$ must be close to the Fermi energy. Such a
circumstance is easily available when the U and L band Fermi surfaces
intersect. Of course, bands repel each other in general unless there
is a  symmetry that forbids mixing. Thus, Cooper pairing will occur in the
vicinity of special symmetry points in the Brillouin zone.  

Our previous paper dealt with the unphysical situation of two distinct
bands with exactly coincident Fermi surfaces at a given doping which
was associated with the optimal doping for superconductivity. In this
regard, the previous study of IBP must be regarded as an overly
simplistic model that serves to illustrate some of the general
principles of IBP but cannot account for quantitative features.
This paper considers realistic bands for the cuprates and derives some
quantitative consequences of the theory. 

We propose the two relevant bands for high T$_c$ are comprised
predominantly of Cu $d_{x^2-y^2}$, Cu 
$d_{z^2}$, planar O $p_\sigma$, and apical O $p_z$ orbitals.
The two bands are
strong mixtures of these orbitals except at special k points only. A
simple Hubbard model is written down for the system. The parameters
used in our Hubbard model are derived in the following paper 
by Perry and Tahir-Kheli$^4$ using
ab-initio calculations on finite clusters for 
La$_{1.85}$Sr$_{0.15}$CuO$_4$. Contrary to the
results of local density approximation (LDA) band structure
calculations on the cuprates, we find two bands at the Fermi energy
comprised primarily of the above orbitals. The difference between our
results and previous band structure calculations are attributed to
correlation effects.
The
optimal doping is the doping where the two band Fermi surfaces
touch. This can happen along the diagonal where $k_x=k_y$. 
La$_{1.85}$Sr$_{0.15}$CuO$_4$ is
primarily considered here so the 3D dispersion is small. 

The NMR spin relaxation rates at the planar Cu and O nuclei are
estimated with these bands and so are the respective Knight
shifts. Reasonable quantitative agreement with many of the anomalous
NMR properties$^{5-11}$
is obtained. The dramatically different temperature
dependencies of the planar Cu and O spin relaxation rates is due to
the conspiring of several important aspects of the orbital characters
of the two bands U and L at the Fermi energy. 
The major reason is that there are two possible orbitals on the Cu,
$d_{x^2-y^2}$ and $d_{z^2}$, whereas there is only one important
orbital $p_\sigma$ on the planar O. The second most important reason
is that the L band is almost full at optimal doping and the orbital
character of the available L k states is dominated by the character at
the top of the band at $k=(\pi,\pi)$.

The large spin
relaxation anisotropy$^{10}$
on the Cu site is discussed and we argue this is
due to a small mixture of $d_{xy}$ character in the two bands. We show
that mixing of $d_{xy}$ of a few percent is sufficient to produce the
observed anisotropy ratio of $\approx3.4$.

The Cu and O Knight shifts are shown to have similar T (temperature)
dependence to the O relaxation rate although strictly speaking, we
expect deviations to exist. Once again, the above mentioned reasons
for the relaxation rates come into play.
Finally, we show
that the extra 3D dispersion one expects for these bands in optimally
doped YBa$_2$Cu$_3$O$_7$ can resolve the lack of T dependence of the
Knight shifts and O relaxation rate observed for this cuprate.

With the same parameters, the Hall effect is calculated using
standard Bloch-Boltzman theory and is shown to have the observed
monotonic decreasing T dependence varying by $\approx50\%$ over the
temperature range compared to the experimental$^{12}$
value of $\approx50\%$, with
absolute values about 13 times larger than measured values.

The linear resistivity is also obtained due to phonon relaxation
across the bands. The sensitivity of R$_h$ and $\rho$ to the doping
follows naturally from the model. The qualitative features of the
above experiments are straightforward from the considerations
presented here. These features are very robust and do not depend
sensitively at all on the choice of parameters used. This is
particularly gratifying because these features are observed for a
variety of different materials. Derivations of so many of the
anomalous normal state properties of the cuprates lends confidence to
the overall theory.

Section II motivates and defines the IBP model. A Hamiltonian is
written down and solved within the BCS mean field approximation. The
orbital character of our bands is chosen. The optimal doping for
superconductivity as the Fermi surface touching point is
discussed. The lack of a unique phase for the attractive coupling
matrix element V$_{kk'}$ is discussed and the difference between
Josephson tunneling in standard BCS-like intra-band pairing and IBP
schemes is derived. IBP due to coupling through phonons is
proposed. Up to this point the arguments and equations presented in
Section II are quite general. There are several different channels
through which IBP could occur. One particular channel is chosen and we
restrict ourselves to singlet pairs only. There is nothing in the
general theory that precludes IBP due to phonons with triplet pairs as
far as we can see. 

The 4 key Josephson tunneling experiments$^{2,3,13,14}$
are discussed. We show that
with our choice of pairing channel, the two ``d-wave'' results of
Wollman$^2$ et al and Tsuei$^3$ 
et al can be understood. The observed c-axis
tunneling of Dynes$^{13}$ et al can also be understood as due to BCS-like
pairs formed away from the Fermi surface touching point. We are unable
at present to decide whether the ``s-wave'' result of Chaudhari and
Lim$^{14}$
supports or contradicts our model due to the complexity of the
single particle tunneling matrix elements in this case.

Section III presents quantitative calculations of the various normal
state properties mentioned above and the final section is devoted to a
summary of our results.

\bigskip
\centerline{\bf II. INTER-BAND PAIRING MODEL}
\smallskip
\noindent{\bf A. The Model}

The T$_c$ of the cuprates is quite sensitive to the doping. Typically,
there exists one optimal doping$^{15}$
for the highest T$_c$ and a rapid
change in T$_c$ above and below this doping. A notable exception is
YBCO$^{16}$
where one may argue that there are two optimal dopings,
T$_c=62$K, $92$K with an intermediate crossover regime. Regardless,
whether we argue for one or two optimal dopings for YBCO, it is
natural to associate the change in T$_c$ with doping to an attractive
coupling that is strongly doping dependent. Experimentally,$^{17}$
many
normal state properties are simultaneously strongly dependent on
the doping. Thus, one is led to propose a bosonic excitation that
couples to the electronic charge where either (or both) the bosons or
their electronic coupling is strongly renormalized by the doping
level. Rather than such dynamic schemes for the change in the
attractive coupling with doping, we ask if there exists a simple
kinematic scheme for the doping dependence of T$_c$. In such a scheme,
we hope to find a pairing strength that weakly depends on doping with
the suppression of T$_c$ due primarily to a kinematic mechanism. IBP is
such a mechanism. 

Consider two bands labeled U and L with dispersions $\ep^{(U)}_k$,
$\ep^{(L)}_k$ and further suppose there exists some attractive coupling
leading to Cooper pairs of the form $(k_U\up,-k_L\down)$ or
$(k_L\up,-k_U\down)$. In general, $\ep^{(U)}_k\neq\ep^{(L)}_k$. It is only
when $\ep^{(U)}_k$ and $\ep^{(L)}_k$ are both close to the Fermi energy that
there is any possibility of a lowering of the overall energy by the
formation of a Cooper pair. In BCS superconductors, pairs are formed
for k states satisfying,
$$|\ep_k - \ep_F| < \hbar\omega_D,\eqno(1)$$
\noindent where $\hbar\omega_D$ is the Debye energy and $\ep_F$ is the
Fermi energy.
For IBP, the
analogous criteria are 
$$|\ep^{(U)}_k-\ep_F| < \hbar\omega_D,\eqno(2)$$
$$|\ep^{(L)}_k-\ep_F| < \hbar\omega_D.\eqno(3)$$
\noindent For BCS superconductors, there are always k states in the
required energy range for pairing regardless of the doping. This is no
longer the case for IBP because (2) and (3) must both be satisfied. It is
only for special dopings that the U band and L band Fermi surfaces
touch or are close to touching that any possibility exists for
IBP. Hence, with IBP the doping sensitivity of T$_c$ can be
understood quite simply as due to the change in the minimal separation
in k space of the U and L band Fermi surfaces. No change in the
strength of the pair attraction is required. In effect, we have
replaced the requirement of a doping dependent pairing interaction by
a geometric argument based upon the band energy differences in k
space.

We will of course, take the mediator of the inter-band attraction to
be phonons as in BCS.  Because our pairs are comprised of electrons
from two bands, the standard arguments for an upper bound of
T$_c\approx30$K do not apply. This is certainly gratifying as it does not
exclude IBP with phonons as a theory for high temperature
superconductors. At present we have no estimate for
the expected range of values for T$_c$ with our mechanism.

There are some important differences between the pairing terms in the
inter-band Hamiltonian and a BCS Hamiltonian. In order to highlight
these differences we will briefly re-derive some of the well known
aspects of the BCS Hamiltonian in order to both generalize to IBP and
to establish our notation.

In general, spin is conserved. Thus, there are two kinds of
scatterings of BCS-like $(k,-k)$ pairs: singlet to singlet, and triplet to
triplet. Let the matrix elements for scattering a $(k,-k)$ singlet
(triplet) to a $(k',-k')$ singlet (triplet),

$$\rt2(\phi_k\phi_{-k}\pm\phi_{-k}\phi_k) \rt2(\up\down\mp\down\up)
\rightarrow
   \rt2(\phi_{k'}\phi_{-k'}\pm\phi_{-k'}\phi_{k'})
      \rt2(\up\down\mp\down\up),\eqno(4)$$

\noindent be $V^0_{k'k}$ $(V^1_{k'k})$.  $\phi_k$ is a single particle
wavefunction of momentum k. The first important point is that although
$\phi_k$ is uniquely defined only up to a phase
$\phi_k\rightarrow\exp^{i\theta(k)}\phi_k$, with
$\theta(-k)=-\theta(k)$, $V^{0,3}_{k'k}$ is uniquely defined due to
the time reversal invariance of a pair with itself.

The Hamiltonian being Hermitian means $V^{0,1}_{k'k}=V^{0,1*}_{kk'}$
and time reversal invariance of $H$ implies
$V^{0,1}_{k'k}=V^{0,1}_{kk'}$. Thus, $V_{k'k}^{0,1}$
is always real. In
second quantized form,

$$\rt2(\phi_k\phi_{-k}\pm\phi_{-k}\phi_k) \rt2(\up\down\mp\down\up)
=
\rt2(a^\dagger_{k\up}a^\dagger_{-k\down}\mp
     a^\dagger_{k\down}a^\dagger_{-k\up}),\eqno(5)$$

\noindent where $a^\dagger_{k\sigma}$ is the creation operator for an
electron in state $\phi_k$ with spin $\sigma$.
The general pair-pair interaction is,

$$\sum_{k'k}V^1_{k'k}\{\o2
(
a^\dagger_{k'\up}a^\dagger_{-k'\down}+
a^\dagger_{k'\down}a^\dagger_{-k'\up}
)
(
a_{-k\down}a_{k\up}+a_{-k\up}a_{k\down}
)\ +$$
$$
a^\dagger_{k'\up}a^\dagger_{-k'\up}a_{-k\up}a_{k\up}
+
a^\dagger_{k'\down}a^\dagger_{-k'\down}a_{-k\down}a_{k\down}\}\ 
+$$
$$ \sum_{k'k}V^0_{k'k}\biggl(\o2\biggr)
(
a^\dagger_{k'\up}a^\dagger_{-k'\down}-
a^\dagger_{k'\down}a^\dagger_{-k'\up}
)
(
a_{-k\down}a_{k\up}-
a_{-k\up}a_{k\down}
),
\eqno(6)$$

\noindent which can be expressed as,

$$\sum_{k'k}V_{k'k}
a^\dagger_{k'\up}a^\dagger_{-k'\down}a_{-k\down}a_{k\up}\ +
\sum_{\scriptstyle k'k\atop\sigma} {1\over4}(V_{k'k}-V_{k'-k})
a^\dagger_{k'\sigma}a^\dagger_{-k'\sigma}
a_{-k\sigma}a_{k\sigma},\eqno(7)$$

\noindent where,

$$V_{k'k}=\o2\bigl[(V^0_{k'k}+V^0_{-k'-k}+V^0_{k'-k}+V^0_{-k'-k})+
(V^1_{k'k}+V^1_{-k'-k}-V^1_{k'-k}-V^1_{-k'k})\bigr].\eqno(8)$$

From the relations for $V_{k'k}^{0,1}$,
we have $V_{k'k}=V_{kk'}$ in general,
and $V_{k'k}=V_{k'-k}$ for pure singlet pairing, and $V_{k'k}=-V_{k'-k}$
for pure triplet pairing. Finally, $V_{k'k}$ is always real and
uniquely defined. For phonon coupling, $V_{k'k}$ is constant leading to
pure singlet pairs. The symmetry of $V_{k'k}$ determines the total
spin of the Cooper pairs in BCS. This is not true for IBP.

For IBP, let the matrix elements for the scatterings,

$$\rt2(\phi_{Uk}\phi_{L-k}\pm\phi_{L-k}\phi_{Uk})
\rt2(\up\down\mp\down\up)
\rightarrow
\rt2(\phi_{Uk'}\phi_{L-k'}\pm\phi_{L-k'}\phi_{Uk'})
\rt2(\up\down\mp\down\up),\eqno(9)$$

\noindent be $V^0_{k'k}$, $V^1_{k'k}$. Here, $\phi_{U,Lk}$ are the single
electron wavefunctions for each band with momentum $k$.
In this case $V_{k'k}^{0,1}$ is no
longer uniquely defined for a change in the definitions of $\phi_{Uk}$
and $\phi_{Lk}$,

$$\phi_{Uk}\rightarrow\exp^{i\theta_U(k)}\phi_{Uk},\eqno(10)$$
$$\phi_{Lk}\rightarrow\exp^{i\theta_L(k)}\phi_{Lk},\eqno(11)$$

\noindent where $\theta_{U,L}(-k)=-\theta_{U,L}(k)$ implies,

$$V_{k'k}^{0,1}\rightarrow\exp^{-i[\theta_U(k')-\theta_L(k')]}
\exp^{i[\theta_U(k)-\theta_L(k)]}V_{k'k}^{0,1},\eqno(12)$$

Hermiticity implies,
$V_{k'k}^{0,1}=V_{kk'}^{0,1*}$ and time reversal symmetry leads to
$V_{k'k}^{0,1}=V_{-k-k'}^{0,1}$ 
rather than $V_{k'k}^{0,1}=V_{kk'}^{0,1}$ for BCS. This is due to
inter-band Cooper pairs not being time reversal invariant with
themselves. 

The general pair-pair inter-band pairing interaction is,

$$\sum_{k'k}V^1_{k'k}\{\o2
(
a^\dagger_{Uk'\up}a^\dagger_{L-k'\down}+
a^\dagger_{Uk'\down}a^\dagger_{L-k'\up}
)
(
a_{L-k\down}a_{Uk\up}+a_{L-k\up}a_{Uk\down}
)\ +$$
$$
a^\dagger_{Uk'\up}a^\dagger_{L-k'\up}a_{L-k\up}a_{Uk\up}
+
a^\dagger_{Uk'\down}a^\dagger_{L-k'\down}a_{L-k\down}a_{Uk\down}\}\ 
+$$
$$ \sum_{k'k}V^0_{k'k}\biggl(\o2\biggr)
(
a^\dagger_{Uk'\up}a^\dagger_{L-k'\down}-
a^\dagger_{Uk'\down}a^\dagger_{L-k'\up}
)
(
a_{L-k\down}a_{Uk\up}-
a_{L-k\up}a_{Uk\down}
),
\eqno(13)$$

\noindent These equations make it clear that one cannot even speak of
the symmetry of the pairing $V_{k'k}$ without first specifying the
phase convention on the single particle orbitals. Secondly, say we
found a phase convention such that $V^{0,1}_{k'k}$ is pure
s-wave. Unlike BCS phonon pairing, IBP does not preclude triplet
pairing due to phonons. Experimentally, pure triplet pairs are not
consistent with the observed Josephson tunneling between a BCS
superconductor and the cuprates. There must be some singlet pairs for
tunneling to occur. This does not exclude the possibility of some
pairs being triplet paired, though. For the rest of this paper however,
we will only consider the case of pure singlet pairs, i.e. $V^1_{k'k}=0$.

The IBP Hamiltonian becomes,

$$H=\sum_{k\sigma}\biggl[
\ep^{(U)}_ka^\dagger_{Uk\sigma}a_{Uk\sigma} + 
\ep^{(L)}_ka^\dagger_{Lk\sigma}a_{Lk\sigma}\biggr]\quad+$$
$$\sum_{k'k}\o2V_{k'k}(a^\dagger_{Uk'\up}a^\dagger_{L-k'\down} - 
a^\dagger_{Uk'\down}a^\dagger_{L-k'\up})
(a_{L-k\down}a_{Uk\up}-a_{L-k\up}a_{Uk\down}),\eqno(14)$$

\noindent where we write $V_{k'k}=V^0_{k'k}$ for simplicity.

In the pairing approximation, we consider excitations of the form,

$$\psi_0(k)
=\left(u_k+v_ka^\dagger_{Uk\up}a^\dagger_{L-k\down}\right)
|0\rangle,\ \ \ \ \ p_0(k)=(1-f_k^{(U)})(1-f_k^{(L)}),\eqno(15)$$
$$\psi_1(k)=a^\dagger_{Uk\up}|0\rangle,\ \ \ \ \
p_1(k)=f_k^{(U)}(1-f_k^{(L)}),\eqno(16)$$
$$\psi_2(k)=a^\dagger_{L-k\down}|0\rangle,\ \ \ \ \ 
p_2(k)=(1-f_k^{(U)})f_k^{(L)},\eqno(17)$$
$$\psi_3(k)=
\left(-v_k+u_ka^\dagger_{Uk\up}a^\dagger_{L-k\down}\right)|0\rangle,
\ \ \ \ \ p_3(k)=f_k^{(U)}f_k^{(L)},\eqno(18)$$
$$|u_k|^2+|v_k|^2=1,\eqno(19)$$

\noindent with probabilities $p_i(k)$ where $f_k^{(U)}$ and
$f_k^{(L)}$ are the occupation numbers of U and L band particles of
momenta $k$ and $-k$ respectively. There are four other excitations
$\psi_4,\ldots,\psi_7$ obtained from the above by replacing
$a^\dagger_{Uk\up}a^\dagger_{L-k\down}$ with
$a^\dagger_{L-k\up}a^\dagger_{Uk\down}$ in $\psi_0$ and $\psi_3$,
$a^\dagger_{Uk\up}$ with $a^\dagger_{L-k\up}$ in $\psi_1$ and
$a^\dagger_{L-k\down}$ with $a^\dagger_{Uk\down}$ in $\psi_2$.
The ordering of the creation operators is chosen such that Cooper
pairs consist only of singlet pairs.

The coefficients $u_k$ and $v_k$ are chosen to minimize the free
energy
$F=H-\mu N-TS$. The solution is,

$$|u_k|^2=1-|v_k|^2={1\over 2}\left(1+{\xi_k\over E_k}\right),\ \ 
u_k^*v_k={\Delta_k\over 2E_k},\eqno(20)$$
$$\xi_k=\omega_k-\mu,\ \ \ \ \ 
\omega_k={1\over 2}\left(\ep_k^{(U)}+\ep_k^{(L)}\right),\eqno(21)$$
$$E_k=\sqrt{\xi_k^2+|\Delta_k|^2},\eqno(22)$$
$$E^{(U)}_k={1\over 2}[\ep^{(U)}_k-\ep^{(L)}_k]+E_k,\eqno(23)$$
$$E^{(L)}_k=-{1\over 2}[\ep^{(U)}_k-\ep^{(L)}_k]+E_k,\eqno(24)$$
$$f_k^{(U)}={1\over \exp^{\beta E_k^{(U)}}+1},\ \ \ \ \ 
f_k^{(L)}={1\over \exp^{\beta E_k^{(L)}}+1},\eqno(25)$$

\noindent with gap equation,

$$\Delta_k=\sum_{k'}V_{kk'}{\Delta_{k'}\over 2E_{k'}}
                     (1-f_{k'}^{(U)}-f_{k'}^{(L)}).\eqno(26)$$

Conservation of the total number of particles $N$ leads to an equation for
the chemical potential $\mu$,

$$N=\sum_k 2|v_k|^2(1-f_k^{(U)}-f_k^{(L)})+
(f_k^{(U)}+f_k^{(L)}),\eqno(27)$$

Unlike BCS,
the quasiparticle excitation energies are different for the U and
L particles and are given by $E_k^{(U)}$ and $E_k^{(L)}$. These
energies are the sum of half the difference in the U and L band
energies and the term $E_k$ which is analogous to the BCS
quasiparticle energy. The most important point of all is that, although
the sum $E_k^{(U)}+E_k^{(L)}=2E_k$ is always positive, the U or L 
excitation energy can be negative or have lower energy than the gap energy
$\Delta$.
This new piece
of physics will be considered in more detail below.

Also, the BCS quasiparticle energy $E_k$ is formed from a band
$\omega_k$ that is the mean of the U and L bands. One can see why this
is the case by noting that in the BCS ground state, pairs are either
fully occupied or completely unoccupied. 
The individual band energies always appear summed together,
$\ep_k^{(U)}+\ep_k^{(L)}=2\omega_k$.
For these states, the system ``doesn't know'' that the U and L pairing
electrons have different energies.
The three equations for the pair occupation amplitudes $u_k$ and
$v_k$, incorporate the difference between the two band dispersions
only through the gap $\Delta_k$. This is to be expected by the same
argument as above because $u_k$ and $v_k$ represent pair occupations.
Similarly, the gap equation incorporates the U and L band differences
through the quasiparticle occupation numbers, $f^{(U)}$ and $f^{(L)}$.

Looking at the expressions for the quasiparticle excitation energies
$E_k^{(U)}$ and $E_k^{(L)}$, one sees that the size of the difference in
energies of the two bands $\ep^{(U)}_k-\ep^{(L)}_k$ is what determines
the size of the excitation. A negative energy implies that no Cooper
pair is formed at $T=0$. Instead, there is a single free electron in
one band and no electron in the other. When $\ep^{(U)}_k-\ep^{(L)}_k$
is large, it is energetically unfavorable to occupy both k states or
empty both k states. Thus no pair is formed. Near a Fermi surface
touching point, $\ep^{(U)}_k\approx\ep^{(L)}_k$. Therefore, 
inter-band pairs are
always energetically favored.

These equations are very similar to the equations for gapless
superconductivity.$^{18}$ In gapless superconductivity, an $\up$ spin
electron has a slightly different dispersion than the $\down$ spin
electrons due to the presence of magnetic impurities. These $\up$ and
$\down$ spin bands are the analogues of the U and L bands in IBP.

\medskip
\noindent {\bf B. The Two Relevant Bands}

Perry and Tahir-Kheli$^4$ have calculated ab-initio the existence of two
bands near the Fermi energy for La$_{1.85}$Sr$_{0.15}$CuO$_4$.
The results of that work are
briefly summarized in this sub-section.

La$_2$CuO$_4$ has two structural phases. A high temperature
body-centered tetragonal crystal with D$_{4h}$ point group and a low
temperature orthorhombic lattice with C$_{2h}$ point group. In the low
temperature phase, the CuO$_6$ octahedra are tipped by $4.3^\circ$
from their high temperature positions thereby reducing the symmetry of
the crystal. The low temperature phase is the structure for
superconductivity. In the tetragonal crystal, there are two reflection
planes defined by the z-axis and the lines $x=y$ and $x=-y$
respectively. For the orthorhombic crystal, there is only one
reflection plane defined by the z-axis and $x=y$.

In either case, there will be a rigorous band crossing along $k_x=k_y$
for our two bands. The tetragonal phase will also have a crossing
along $k_x=-k_y$. The orthorhombic phase will come close to crossing
along $k_x=-k_y$, but cannot cross. For IBP to occur, it is imperative
that a crossing exist, for otherwise it is hard to see how IBP can
overcome the band repulsion. For the remainder of this paper and in
the following paper, we will take the LASCO crystal structure to be
the high temperature tetragonal phase for simplicity. The small
difference in structures can have a big effect on T$_c$, but for most
normal state properties, the difference will be small. 

In undoped La$_2$CuO$_4$, the O sites have a formal charge of
$-2$, the Cu charge is $+2$ and is in a d$^9$ state. The point charge
field on the Cu site due to the surrounding atoms make the
$d_{x^2-y^2}$ orbital the most unstable leading to holes in this
orbital in the undoped system. The next highest (unstable) orbital on
the Cu is $d_{z^2}$ due to the field of the apical oxygens. Undoped,
each Cu $d_{x^2-y^2}$ has a single electron and $d_{z^2}$ has 2
electrons. We propose that as the system is doped to its
metallic/superconducting phase, holes start to appear in the Cu
$d_{z^2}$ orbital. At an arbitrary k point in the Brillouin zone,
$d_{z^2}$ can mix with $d_{x^2-y^2}$, O $p_\sigma$, and apical O $p_z$
orbitals. Thus,
the two relevant bands at the Fermi energy are comprised
primarily of these four orbitals.

This assumption of the existence of $d_{z^2}$ holes near the Fermi
energy along with the assumption of inter-band pairing must be
regarded as the two most important postulates of the IBP model for the
cuprates. The former postulate is discussed in detail using ab-initio
calculations on small clusters in the following paper$^4$ where we
conclude that there are two bands near the Fermi energy with the
character described above. 

We will call the two bands the Upper (U) and Lower (L) bands where
$\ep^{(U)}_k\ge\ep^{(L)}_k$. The lower band should be full or almost
full in the undoped case and the upper band is half full with all of
its holes predominantly of $d_{x^2-y^2}$ character. For LaSrCuO, the U
and L bands will be almost completely 2D, while for optimally doped
YBa$_2$Cu$_3$O$_7$, the bands will have a measurable amount of 3D
character. Underdoped
YBa$_2$Cu$_3$O$_{6.63}$ will have less 3D dispersion than optimally
doped YBa$_2$Cu$_3$O$_7$.
Additionally, for 
YBCO there are potentially 4 relevant bands due to
the 2 CuO planes per unit cell. Here, we restrict our attention to
LaSrCuO and briefly discuss the differences to expect for YBCO where
appropriate. 

The unit cell of two formula units of La$_2$CuO$_4$ is tetragonal with
lattice spacing $a=4.0$~\AA\ in the $x$, $y$ direction and $c=12.0$ 
\AA\ in the z-direction. The Brillouin zone of the primitive unit cell is
given by $-\pi/a\le k_x\le\pi/a$, $-\pi/a\le k_y\le\pi/a$,
$-2\pi/c\le k_z\le2\pi/c$.

To a first approximation, we may take the U and L bands for LaSrCuO to
be purely 2D and add in the weak z-axis dispersion as a first order
perturbation. This is done primarily for further
computational simplicity and to convey the key aspects of the model. A
more correct description will not qualitatively alter the 
behaviors we
obtain for the NMR and Hall effect and more importantly, will not
affect the general arguments for the various normal state properties.

The relevant orbitals are $d_{z^2}$, $d_{x^2-y^2}$ on the Cu,
$p_\sigma$ orbitals on the two planar O sites ($\sigma$ along the CuO
bond direction) and $p_z$ orbitals for the two apical O sites above
and below the Cu atom. Additional orbitals that appear are the Cu
$4s$, $d_{xy}$, and O $p_\pi$ in the plane and the $p_x$, $p_y$
orbitals on the apical O's. None of these additional orbitals will
lead to a big change in the band structure but can affect the nmr. In
particular, a small amount of Cu $4s$ is required for an understanding
of the sign of the Knight shift on Cu and some (a few percent)
$d_{xy}$ is necessary for explaining the large anisotropy in the Cu
spin relaxation rates for planar and z-axis magnetic fields.

The tight-binding Hubbard model is (we use the electron picture),

$$H=H_{\rm orb} + H_{\rm hop},\eqno(28)$$

$$H_{\rm orb}=\sum_{n\sigma}
\ep_{d_{x^2-y^2}}d^\dagger_{x^2-y^2n\sigma}d_{x^2-y^2n\sigma} +
\ep_{d_{z^2}}d^\dagger_{z^2n\sigma}d_{z^2n\sigma}\quad+$$

$$\sum_{n\sigma}
\ep_{p_\sigma}(p^\dagger_{xn\sigma}p_{xn\sigma}+
p^\dagger_{yn\sigma}p_{yn\sigma}) + 
\ep_{p_z}(p^\dagger_{Uzn\sigma}p_{Uzn\sigma} +
p^\dagger_{Lzn\sigma}p_{Lzn\sigma}),\eqno(29)$$

$$H_{\rm hop}=\pm t_{x^2-y^2,\sigma}
\sum_{{\langle nm\rangle\atop\sigma}}
d^\dagger_{x^2-y^2n\sigma}(p_{xm\sigma} + p_{ym\sigma})
\pm t_{z^2,\sigma}\sum_{{\langle nm\rangle\atop\sigma}}
d^\dagger_{z^2n\sigma}(p_{xm\sigma} + p_{ym\sigma})\quad+$$

$$\pm t_{\sigma\sigma}
\sum_{<nm>\sigma}p^\dagger_{xn\sigma}p_{ym\sigma}
\pm t_{p_z,d_{z^2}}\sum_{n\sigma}
d^\dagger_{z^2n\sigma}(p_{zUn\sigma}-p_{zLn\sigma}) + 
t_{p_z,p_z}\sum_{n\sigma} p^\dagger_{zUn\sigma}p_{zLn\sigma}\ +$$

$$\pm t_{\sigma\sigma a}\sum_{\langle nm\rangle\sigma\atop
\rm{same\atop axis}}
(p^\dagger_{xn\sigma}p_{xm\sigma}+p^\dagger_{yn\sigma}p_{ym\sigma})
\pm t_{p_z,\sigma}\sum_{\langle nm\rangle\atop\sigma}
(p^\dagger_{zUn\sigma}-p^\dagger_{zLn\sigma})
(p_{xm\sigma}-p_{ym\sigma})+{\rm c.c.},\eqno(30)$$

\noindent where $d_{z^2n\sigma}^\dagger$ 
creates a $d_{z^2}$ orbital with spin
$\sigma$ at site $n$. $p^\dagger_{xn\sigma}$ ($p^\dagger_{yn\sigma}$)
creates a $p_\sigma$ electron on a planar O site along the $x$ ($y$)
axis and $p_{zUn\sigma}^\dagger$ ($p_{zLn\sigma}^\dagger$) creates a
$p_z$ electron on the apical O site above (below) the CuO plane.
$\ep_{d_{z^2}}$ is the self (orbital) energy of the
$d_{z^2}$, etc. The t's are the various hopping matrix elements.
The $\pm$ sign in front of the hopping matrix
elements represents the fact that the sign of the matrix element
depends on the relative position of the two relevant orbitals.
$\langle nm\rangle$ represents neighboring sites.
The values of the parameters are shown in Table 1. The values in Table
1 for the hopping t's is for the antibonding combination of the two
relevant orbitals.

Except for the value of the difference
$\ep_{d_{z^2}}-\ep_{d_{x^2-y^2}}$, there is nothing particularly
surprising or out of the ordinary with these parameters. 
$\ep_{d_{z^2}}-\ep_{d_{x^2-y^2}}$ being positive is what brings the
$d_{z^2}$ orbital up to the Fermi energy. This is the essential
parameter for a two band description like ours of
superconductivity. These points are discussed in detail in the
following paper$^4$ where our parameters are derived from ab-initio
calculations on small clusters.
As we stated in the introduction, the theory presented here is very
robust and at a qualitative level does not depend 
sensitively on the values for
these parameters at all. 

The point group of LaSrCuO is D$_{4h}$ and $d_{x^2-y^2}$, $d_{z^2}$
transform as B$_{1g}$ and A$_{1g}$ respectively. Under a $\sigma_d$
reflection about the diagonals $x=\pm y$,
$\sigma_dd_{x^2-y^2}=-d_{x^2-y^2}$ and
$\sigma_dd_{z^2}=d_{z^2}$. Thus, for diagonal k vectors $(k_x=\pm k_y)$
a single electron wavefunction must have $d_{x^2-y^2}$ character or $d_{z^2}$
character. Along $k_x=\pm k_y$, two bands may cross if one band has
$d_{x^2-y^2}$ character and the other has $d_{z^2}$ character. A plot
of the top two bands (most unstable) is shown in figure 1 along the
closed path $(0,0)-(\pi,\pi)-(\pi,0)-(0,0)$ in $k$ space.
For all k points not
on the diagonal, there is no symmetry to keep $d_{x^2-y^2}$ and
$d_{z^2}$ from mixing. Thus, the two bands will repel leading to
$\ep^{(U)}_k>\ep^{(L)}_k$. 

The optimal doping for the highest T$_c$ is the doping that leads to a
Fermi energy equal to the energy at the band crossing or touching
point. It is at this doping and this doping only that there is
favorable energetics for inter-band pair formation. Figure $2$ shows the
Fermi surface at various doping from underdoped to overdoped. 

We have adjusted the charge transfer of planar O $p_\pi$ onto the La
described in detail in the following paper$^4$
such that this doping matches the
experimentally observed optimal doping for La$_{2-x}$Sr$_x$CuO$_4$ at
$x=0.15$. Undoped, there are a total of 3
electrons in the U and L bands while for $x=0.15$ there are
$3-0.15=2.85$ electrons in the two bands.

The z-axis (normal to the CuO planes) 
dispersion is approximately calculated by observing the
dominant coupling in the z-direction is the $p_z$ orbital on the
apical O above a CuO plane coupling to the $p_z$ orbital on the apical
O below the next higher CuO plane. We take this value to be
$T_{(0,0)}=0.3$ eV. The coupling is added in via first order
perturbation theory. Our choice for this matrix element is described
below. 

The additional energy of a k state due to $p_z$ to $p_z$
coupling from layer to layer is,

$$\ep(k_x,k_y,k_z)=-2{\rm T}_{(0,0)}{\rm P}\cos\o2 (k_xa)
\cos\o2 (k_ya)\cos
k_zc,\eqno(31)$$ 

\noindent where P is the amount of apical O $p_z$ character in the k
state $(k_x,k_y)$ derived from the 2D Hamiltonian (28).

To incorporate the very small couplings through the $p_x$ and $p_y$
orbitals on the apical O above a CuO plane with the $p_z$ on the
apical O below the next higher plane,
an additional energy,

$$\ep(k_x,k_y,k_z)=-2{\rm T}_{(\pi,0)}{\rm P}
\bigl[\cos\o2 (k_xa)\sin\o2 (k_ya) + \sin\o2 (k_xa)\cos\o2 (k_ya)\bigr]
\cos k_zc,\eqno(32)$$

\noindent is added. Finally, we add,

$$\ep(k_x,k_y,k_z)=-2{\rm T}_{(\pi,\pi)}{\rm P}
\sin\o2 (k_xa)\sin\o2 (k_ya)\cos k_zc,\eqno(33)$$

\noindent to include the weak coupling of Cu $d_{xy}$ with apical O $p_z$.
We take T$_{(\pi,0)}=0.05$ eV, T$_{(\pi,\pi)}=0.02$ eV. 

The reason for naming the hopping matrix elements with the k space
labels $(0,0)$, $(\pi,0)$, $(\pi,\pi)$ is that each one is multiplied
by a combination of $\cos\o2 k_xa$, $\cos\o2 k_ya$, $\sin\o2 k_xa$,
$\sin\o2 k_ya$ which is equal to 1 on the particular k space label and
zero at the other two. We expect 
T$_{(0,0)} >{\rm T}_{(\pi,0)}>{\rm T}_{(\pi,\pi)}$. 

The primary effect of adding in the above z-axis couplings is to
eliminate the logarithmic 2D van-Hove saddle point
singularity in the density of
states of the U band
by a broadened 3D peak. The width of this peak is responsible
for the T dependence (or lack thereof in YBa$_2$Cu$_3$O$_7$) of the Knight
shifts in the normal state. This is shown further along in the
calculation of the nmr.

Figure 3 shows the density of states of the two bands with the Fermi
energy at the optimal doping indicated.
The large peak in the U band density of
states just above the 
optimal doping Fermi energy $\ep_{F}=0.0$ eV is due to the
saddle point singularity at $(\pi,0)$ and $(0,\pi)$. 
The closeness of this peak to $\ep_{F}$ is a
robust feature of this model and is not sensitive to the choice of
parameters. 

Three other features of interest are shown in figures $4(a-c)$. They show the
average amount of $d_{x^2-y^2}$, $d_{z^2}$, and $p_\sigma$ 
character of the two
bands at different energies. Near $\ep_{F}$, the two bands are
predominantly $d_{z^2}$.
Both of these features are once again robust.
\medskip

\noindent {\bf C. The Pairing Term}

There are several different choices for the detailed orbital coupling
that can lead to IBP. Lacking a rigorous microscopic proof that IBP
exists, we make a particular choice for the pairing term and compare
it to experiment. We take the pairing term in equation (14) to be
mediated by an attractive coupling of the form shown in figure 5. In
this figure, $d_{x^2-y^2{\bf k}}$ and $d_{z^2{\bf k}}$ are defined as,

$$d_{x^2-y^2{\bf k}}({\bf r})={1\over\sqrt N}\sum
\exp^{ik\cdot R}d_{x^2-y^2}({\bf r}-{\bf R}),\eqno(34)$$

$$d_{z^2{\bf k}}({\bf r})={1\over\sqrt N}\sum
\exp^{ik\cdot R}d_{z^2}({\bf r}-{\bf R}),\eqno(35)$$

\noindent where {\bf R} is the position of the Cu atom. Eventually, we
will take the origin to be a center of inversion. Figure 5 depicts a
$d_{x^2-y^2}$ Bloch orbital of momentum $-k$ emitting a virtual phonon
and scattering to a $d_{z^2}$ orbital of momentum $-k'$ and a $d_{z^2}$
orbital of momentum $k$ absorbing the phonon and scattering to a
$d_{x^2-y^2}$ orbital of momentum $k'$. (34) and (35) define the phase
convention of the $d_{x^2-y^2}$ and $d_{z^2}$ Bloch functions and the
matrix element U$_{k'k}$. The coupling V$_{k'k}$ in (14) is calculated
by projecting onto the above $d_{x^2-y^2}$, $d_{z^2}$ scattering.
Physically, what is accomplished is quite
straightforward. The attractive coupling between two single electron
states $k$, $-k$ in the same band or two different bands is due primarily
to the attractive coupling of the $d_{x^2-y^2}$ orbital component of
one electron with the $d_{z^2}$ component of the other electron. One
reason that such a coupling may be dominant is due to the
$d_{x^2-y^2}$ $k$ states localizing the electron charge in the plane,
whereas the $d_{z^2}$ $-k$ state localizes the charge out of the CuO
plane thereby reducing the Coulomb repulsion.

An attractive coupling mediated by projection onto the diagram in
figure 5 is also possible for $(k\up,-k\down)$ pairs in the 
{\it same} band
leading to
a traditional BCS pairing term. Thus, one is faced with the
possibility that BCS-like intra-band pairing may win out over
IBP. This possibility is not realized for k states near the band
crossing point
by the same kinematics that
allows a band crossing to occur along the diagonals $(k_x=\pm k_y)$. In
order to have a large BCS-like pairing for a Cooper pair
$(k\up,-k\down)$ in the same band with a k state near the crossing
point one requires a substantial amount of both $d_{x^2-y^2}$ and
$d_{z^2}$ in the k state. Near the band crossing point, the
approximate $\sigma_d$ diagonal reflection symmetry precludes that
leading to a suppression of BCS-like pairing by the above pairing
mechanism. Also, because the U and L k states are almost degenerate
here, the system may further lower its energy by forming a pure
$d_{x^2-y^2}$ and a pure $d_{z^2}$ state in order to maximize the
pairing coupling. At k points away from the diagonals, BCS-like
pairing can occur and for arbitrary dopings, too.

Including BCS-like intra-band pairing by projection of figure 5 as we
did for IBP is straightforward. Rather than diagonalizing a 2x2 matrix
as we did in deriving $(15)-(27)$, we must diagonalize a 4x4
matrix. This is not difficult to do and leads to more complicated
quasiparticle energies than (23) and (24). To properly include
BCS-like pairs, we should not restrict ourselves to the single diagram
in figure 5. There are other combinations of inter-band to BCS-like
pair scatterings that may become important.

Physically, what occurs is near the band crossing point on the
diagonals, inter-band pairs are formed because one band is almost pure
$d_{x^2-y^2}$ (with some O character) and the other band is almost
pure $d_{z^2}$ (again with O character). 
Away from the Fermi surface touching point, only BCS-like
pairs are kinematically allowed leading to standard BCS pairs for the
k states. Thus, one would expect to observe a gap in the photoemission
for k vectors along the $k_x=0$ and $k_y=0$ directions.
This is observed$^{19}$ for angle resolved
photoemission (ARPES)
on Bismuth 2212 although the observed Fermi surface
crossing along $\Gamma M$ ($(0,0)-(\pi,0)$) has been questioned by
Campuzano$^{20}$ et al as due to a k vector shifted by the superlattice
translation $(0.21\pi,0.21\pi)$. 

We also expect a crossover region in
the Brillouin zone where inter-band pairs change over to BCS-like pairs.
Because IBP is only strong at the crossing point along $k_x=k_y$, the
crossover region should occur close to the diagonals. For BCS-like
pairs $(k,-k)$, one would see a gap $\Delta$ in ARPES at this $k$
vector. On the diagonal at the touching point, one would again see a
gap from (23) and (24). Slightly away from the diagonal while there
are still inter-band pairs, (23) and (24) are still valid and at least
one of these excitations has energy less than the gap. Thus, ARPES
should see the ``smallest gap'' slightly off the diagonals $k_x=\pm
k_y$ rather than on the diagonals. We have not yet calculated the size
of this effect to determine whether present experimental resolution
can see this effect.

Owing to the separation of the pairs along the diagonal into one band
with no $d_{x^2-y^2}$ character and the other with no $d_{z^2}$
character, pairs are most strongly formed here than at any other k
point. This suggests that the dominant contribution to the Josephson
tunneling current will come from IBP along the diagonals and we assume
this is true unless such a
current is rigorously zero. One can foresee that such a symmetry
argument will be applied to inter-band tunneling in the c-axis direction
making the contribution zero, thereby obtaining the ``s-wave'' result
of Dynes$^{13}$ et al.

Let,

$$\phi_{Uk}=A_{Uk}d_{x^2-y^2k} + B_{Uk}d_{z^2k} + {\rm other\ terms},
\eqno(36)$$
$$\phi_{Lk}=A_{Lk}d_{x^2-y^2k} + B_{Lk}d_{z^2k} + {\rm other\ terms},
\eqno(37)$$

\noindent where $\phi_{Uk}$ and $\phi_{Lk}$ are the band wavefunctions
and A$_k$, $B_k$ are the projections onto the Bloch functions in
(34), (35). Then,

$$a^\dagger_{Uk}=A_{Uk}d_{x^2-y^2k}^\dagger +
B_{Uk}d^\dagger_{z^2k} + \cdots,\eqno(38)$$
$$a^\dagger_{Lk}=A_{Lk}d_{x^2-y^2k}^\dagger +
B_{Lk}d^\dagger_{z^2k} + \cdots,\eqno(39)$$

\noindent and

$$d_{x^2-y^2k}^\dagger=A_{Uk}^*a_{Uk}^\dagger +
A_{Lk}^*a_{Lk}^\dagger + \cdots,\eqno(40)$$
$$d_{z^2k}^\dagger=B_{Uk}^*a_{Uk}^\dagger +
B_{Lk}^*a_{Lk}^\dagger + \cdots.\eqno(41)$$

The pairing Hamiltonian term is,

$$U_{k'k}(d_{x^2-y^2k'\up}^\dagger d_{z^2-k'\down}^\dagger
d_{x^2-y^2-k\down}d_{z^2k\up} +
d_{x^2-y^2k'\down}^\dagger d_{z^2-k'\up}^\dagger
d_{x^2-y^2-k\up}d_{z^2k\down}).\eqno(42)$$

\noindent Projecting onto (14),

$$V_{k'k}=(A_{Uk'}B_{L-k'})^*(A_{Uk}B_{L-k})U_{k'-k} + 
(B_{Uk'}A_{L-k'})^*(B_{Uk}A_{L-k})U_{-k'k}\ +$$
$$(A_{Uk'}B_{L-k'})^*(B_{Uk}A_{L-k})U_{k'k} +
(B_{Uk'}A_{L-k'})^*(A_{Uk}B_{L-k})U_{-k'-k}.\eqno(43)$$

\noindent Under the phase change in (10) and (11),

$$A_{U,Lk}\rightarrow\exp^{i\Theta_{U,L}(k)}A_{U,Lk},\eqno(44)$$
$$B_{U,Lk}\rightarrow\exp^{i\Theta_{U,L}(k)}B_{U,Lk},\eqno(45)$$

\noindent and thus V$_{k'k}$ transforms as (12).

Now every cuprate space group includes the inversion operator. We
define the origin from which {\bf R} in (34) and (35) is defined as a
point of inversion. With this choice it is easy to see that the
coefficients A$_{U,Lk}$, B$_{U,Lk}$ must always satisfy
A$_k^*$B$_k=$ real.

In general, the electron-phonon matrix element that appears at each
vertex in the Feynman diagram in figure 5 is the sum of terms of the
form,

$$i({\bf e}_{\lambda q}\cdot {\bf q})V_{\lambda q}
(c_{q\lambda} + c_{-q\lambda}^\dagger)
\int\exp^{iq\cdot r}
d_{x^2-y^2k'}^*(r)d_{z^2k}(r)d\tau,\eqno(46)$$

\noindent where $\lambda$ is the phonon polarization and {\bf q} is
its momentum. $c_{q\lambda}$ is the destruction operator and
$V_{\lambda q}$ is the Fourier transform of the phonon potential and
${\bf e}_{\lambda q}$ is the polarization vector. Inversion symmetry
guarantees $V_{\lambda-q}=V_{\lambda q}$, $V_{\lambda q}$ is always
real and also that the integral in (46) over the electron states is
real.

Using (34) and (35), the integral in (46) is,

$$\int\exp^{iq\cdot r}
d_{x^2-y^2k'}^*(r)d_{z^2k}(r)d\tau=
\int\exp^{iq\cdot r}
\sum_{\bf RR'}\exp^{-ik'\cdot R'}\exp^{ik\cdot R}d_{x^2-y^2}(r-R')
d_{z^2}(r-R).\eqno(47)$$

\noindent Taking the largest contribution to be when $R'=R$,

$$\int\exp^{iq\cdot r}
d_{x^2-y^2k'}^*(r)d_{z^2k}(r)d\tau=
\int\exp^{iq\cdot r}d_{x^2-y^2}(r)d_{z^2}(r)d\tau.\eqno(48)$$

Plugging this back into (46), we see that the product of the two
electron-phonon matrix elements due to the two vertices in figure 5
is always
positive with the phase convention defined in (34), (35) where the
origin of R is an inversion center. 
With BCS-like pairing, time reversal symmetry of the Cooper pairs is
sufficient to guarantee the product of the two vertices is always
mod-squared and thus positive.

Therefore, with our definitions of the $d_{x^2-y^2}$, $d_{z^2}$ Bloch
functions, we may take the total coupling $U_{k'k}$ from figure 5 to
be s-like,

$$U_{k'k}=
\cases{-V,&$|\ep^{(U,L)}_k-\mu|<\hbar\omega_D$ and 
$|\ep^{(U,L)}_{k'}-\mu|<\hbar\omega_D$,\cr
\hfill 0,&\hfil otherwise,\hfil\cr}\eqno(49)$$

\noindent leading to,

$$V_{k'k}=(-V)(A_{Uk'}B_{L-k'}+B_{Uk'}A_{L-k'})^*
(A_{Uk}B_{L-k}+B_{Uk}A_{L-k}).\eqno(50)$$

The reason we have chosen the above range of $k'$, $k$ values for
non-zero pairing is seen by considering the relevant Feynman diagram
in figure 5. The value is,

$$\leftline{\indent$\displaystyle
\o2\left[
{1\over{[\ep^{(U)}_{k'}-\ep^{(L)}_k]^2-\hbar^2\omega^2}}+
{1\over{[\ep^{(L)}_{k'}-\ep^{(U)}_k]^2-\hbar^2\omega^2}}\right]\cdot$}$$
$$\hfil(A_{Uk'}B_{L-k'} + B_{Uk'}A_{L-k'})^*
(A_{Uk}B_{L-k} + B_{Uk}A_{L-k})M^2,\eqno(51)$$

\noindent where $M^2$ is a real positive number. If the constraints in
(49) are satisfied,
then the term in brackets is negative. 
\medskip

\noindent{\bf D. Inter-Band Josephson Tunneling}

In BCS theory, a Cooper pair is of the form
$\phi_{k\up}\phi_{-k\down}$ and transforms into itself under the
operation of time reversal. For pairing across two distinct bands as
we propose, the pair $\phi_{Uk\up}\phi_{L-k\down}$ transforms into
$\phi_{Lk\up}\phi_{U-k\down}$ and 
not into itself under time reversal. Of
course, the full Hamiltonian remains time reversal invariant.
This key difference
alters the standard Josephson tunneling in a subtle yet dramatic way
that leads to a new interpretation of the macroscopic phase of
superconductors. 

Consider first the case of Josephson tunneling of a BCS-like 
Cooper pair on one side of a junction to a BCS-like
pair on the other side of the junction.
We will neglect all factors in the expression
for the supercurrent $J$ contributing only to the magnitude and not
to the phase of $J$.
The phase of the pair tunneling 
matrix element for the transfer of a
$(k\up,-k\down)$ pair to a $(p\up,-p\down)$ pair is contained in
the product of two
single-electron tunneling
matrix elements, $T_{kp}$ and $T_{-k-p}$ and the product
of the two gap functions, $\Delta_k^*$ and $\Delta'_p$ on either side
of the junction. This leads to supercurrent,$^{18}$

$$J\propto
T_{kp}T_{-k,-p}\Delta_k\Delta^{'*}_p=|T_{kp}|^2\Delta_k\Delta^{'*}_p.
\eqno(52)$$

\noindent In this expression,
$T_{kp}$ is the matrix element for the transfer of a
single electron of momentum $k$ to an electron of momentum $p$, $T_{-k-p}$
is the corresponding matrix element for transferring the $-k$ electron
to $-p$
and $\Delta_k$, $\Delta'_p$ are the gap functions on
the two sides of the junction. By overall time reversal symmetry,
$T_{-k-p}=T_{kp}^*$. 
Thus, the supercurrent 
is completely controlled by the phases of the
superconducting gap functions on each side of the junction.
The phases of the gap
functions $\Delta, \Delta'$ are determined by the symmetry of the
pairing interactions
for each superconductor. Thus, for all
BCS-like pairing models, Josephson tunneling gives direct
information of the symmetry of the gap.

For IBP, the situation is dramatically different.
Suppose we tunnel from an inter-band superconductor to a BCS
superconductor. 
With Cooper pairing across bands, the matrix element for transferring
a momentum
$k$ electron in the U band to a momentum
$p$ electron on the other side of the
junction, $T_{kp}^{(U)}$, is different from the matrix element
for transferring a $-k$ electron in the L band to a $-p$ electron on
the other side of the junction, $T_{-k,-p}^{(L)}$. 
Although, $T_{-k,-p}^{(U)}=T_{kp}^{(U)*}$ and
$T_{-k,-p}^{(L)}=T^{(L)*}_{kp}$ 
by overall time reversal symmetry, the phase part of
the pair transfer matrix element is of the form,

$$J\propto T_{kp}^{(U)}T_{-k,-p}^{(L)}\Delta_k\Delta^{'*}_p.\eqno(53)$$

\noindent In the case of inter-band pairing,
{\it the symmetry of the pairing interactions and
the orbital character
of the band wavefunctions both contribute to the 
overall tunneling phase}. Under a redefinition of the single band
wavefunctions (10), (11),

$$T^{(U,L)}_{kp}\rightarrow\exp^{i\Theta_{U,L}(k)}T^{(U,L)}_{kp},\eqno(54)$$

\noindent and from the gap equation (26),

$$\Delta_k\rightarrow\exp^{-i[\Theta_U(k)-\Theta_L(k)]}
\Delta_k,\eqno(55)$$

\noindent leading to no change in $J$.
Note also that the above result is independent of
the phase convention of the band orbitals on the BCS side of the
junction because if an electron of momentum $p$ is multiplied by a
phase factor $\exp^{i\theta}$, then the phase of the $-p$ electron is
multiplied by $\exp^{-i\theta}$. 

Remember that as we argued in the previous section, we take the dominant
supercurrent to be due to the inter-band Cooper pairs near the
diagonal crossing points unless the contribution is rigorously zero.

Let us apply our new relation (53) for the supercurrent to the four key
Josephson tunneling experiments on YBCO. The first one by Wollman$^2$
et
al on a YBCO-Pb corner junction is the simplest. We have shown that
the phase of the supercurrent is independent of the choice of single
particle wavefunctions. In this experiment, tunneling occurs along the
two perpendicular Cu-O bond directions in the CuO planes of YBCO (x
and y axes) which are connected by a Pb wire. Pb is a BCS s-wave
superconductor. A phase difference of $\pi$ implies a d-wave gap,
whereas no phase difference implies an s-wave gap. A d-wave result is
obtained. 

Let $T^{(U,L)}_{kp}(x)$ be the matrix element for tunneling a k
electron in the U or L band to a p electron in Pb along the
x-axis. Choose the phase convention on the k states such that a
$90^\circ$ ($C_4$) rotation of the wavefunction for k is equal to the
wavefunction for momentum $C_4k$. Then, we must have
$T^{(U,L)}_{C_4k,C_4p}(y)=T^{(U,L)}_{kp}(x)$. From (36) and (37), our
phase convention gives,

$$A_{U,L}(C_4k)=-A_{U,L}(k),\eqno(56)$$
$$B_{U,L}(C_4k)=+B_{U,L}(k).\eqno(57)$$

Using the gap equation and the expression (50) for $V_{k'k}$,

$$\Delta_k=\Delta(A_{Uk}B_{L-k}+B_{Uk}A_{L-k})^*.\eqno(58)$$

\noindent Hence, 

$$\Delta_{C_4k}=-\Delta_k,\eqno(59)$$

\noindent leading to the result,

$$J(C_4k\rightarrow C_4p)=-J(k\rightarrow p).\eqno(60)$$

\noindent Therefore, our IBP model also gives an observed ``d-wave''
gap.

For the tri-crystal experiment of Tsuei$^3$ et al, we use the fact that for
k near the Fermi surface touching point, $\phi_{Uk}$ has
almost no $d_{z^2}$ character and $\phi_{Lk}$ has almost no
$d_{x^2-y^2}$ character or vice versa. For such a $k$ vector,
$B_{Uk}\approx0$, $A_{Lk}\approx0$ (or $A_{Uk}\approx0, B_{Lk}\approx0$)
and we can take the phase convention $A_{Uk}$, $B_{Lk}$ real with
$A_{Uk}>0$,
$B_{Lk}>0$ (or $B_{Uk}>0, A_{Lk}>0$).
With this convention, $\Delta_k$ is real and $\Delta_k>0$.

Suppose a $d_{x^2-y^2}$ orbital on one side of the grain boundary
tunnels predominantly into a $d_{x^2-y^2}$ state on the other side of
the boundary and thus $d_{z^2}$ on one side goes predominantly
to $d_{z^2}$ on
the other side. Then, the matrix element for $d_{x^2-y^2}\rightarrow
d_{x^2-y^2}$ is proportional to $\cos2\theta\cos2\phi$ where $\theta$
and $\phi$ are the orientations of the x and y axes in the CuO planes
with respect to the grain boundary. Similarly, the $d_{z^2}\rightarrow
d_{z^2}$ matrix element has no orientation dependence and is thus
constant. 

Suppose instead, that the dominant tunneling is
$d_{x^2-y^2}\rightarrow d_{z^2}$ and $d_{z^2}\rightarrow d_{x^2-y^2}$.
In this case, one matrix element is
proportional to $\cos2\theta$ and the other is proportional to
$\cos2\phi$ leading the pair tunneling product $\cos2\theta\cos2\phi$
as before.

Taking the product of the various factors in (53), we see that the
phase controlling the current is the same as to be expected from a
$d_{x^2-y^2}$ gap as is observed.

For the c-axis YBCO-Pb tunneling of Dynes$^{13}$ et al, an analysis similar
to the YBCO corner junction case shows that the inter-band pairs do
not contribute to the current. We still expect to see current due to
the BCS-like pairs in this case.

For the hexagonal YBCO tunneling of Lim$^{14}$ et al, the situation is
different from the tri-crystal YBCO in one very important manner. Here a
hexagonal MgO layer is placed on the LaAlO$_3$ substrate and then YBCO is
grown onto the sample. The hexagonal MgO causes YBCO to grow above it
with its planar x, y axes rotated $45^\circ$ from the angle of the
YBCO grown directly over the LaAlO$_3$ substrate. Thus, we expect a
misalignment of the CuO planes across the six junctions complicating
the matrix elements for single particle tunneling. Unfortunately, we
have been unable to find a convincing argument telling us whether
inter-band or BCS-like pairs dominate the single electron matrix
elements in this case.
\bigskip
\centerline{\bf III. NORMAL STATE NMR AND TRANSPORT}
\medskip
\noindent{\bf A. NMR}

Armed with our parameters for the relevant bands and the criteria
derived from IBP that the optimal doping is when the two Fermi
surfaces touch, it is very easy to compute the normal state nmr and
transport properties using standard expressions for the nmr and the
Bloch-Boltzmann equation for the transport. It is quite satisfying that
not only can we understand qualitatively the numerous anomalous
features of the nmr and transport as due to the character of the bands
at this very special doping, but quantitatively the numbers are
respectable. 

The key normal state nmr features$^{5-11}$
a theory for the cuprates must
explain are: 1.) the difference in the relaxation rate curves with
temperature T for the Cu and O nuclei in the plane, 2.) the similar
Knight shifts (KS) 
at the two sites, 3.) the similarity of the O relaxation
rate over T and the KS, 4.) the lack of T dependence of the KS
for optimally doped YBCO$_7$ and a monotonic increasing T
dependence of the KS for optimally doped LaSrCuO and underdoped
YBCO$_{6.63}$, 5.) the strictly monotonic decreasing Cu relaxation
rate over T for optimally doped YBCO$_7$ and the initially increasing
and then decreasing Cu relaxation rate over T for underdoped
YBCO$_{6.63}$ and optimally doped LaSrCuO, and 6.) the large anisotropy of
the Cu relaxation rate for magnetic fields in the plane and along the
z-axis.

Figure 2 shows the two band Fermi surfaces at optimal doping. The
electron-like band centered at $k=(0,0)$ is the U(pper) band and the
hole-like surface centered around $(\pi,\pi)$ is the L(ower)
band. Both bands are occupied at $k=(0,0)$ and unoccupied at
$(\pi,\pi)$.

Figure 3 shows the density of states for the two bands.
The first thing to notice is that at the Fermi energy
$\ep_F$, the L band has the lower density of states. We expect that
this band predominantly carries the current and being hole-like will
give a Hall coefficient with the correct sign. The large peak in the
density of states of the U band at an energy a little larger ($\approx
0.06$ eV) than $\ep_F$ is due to the saddle point singularity at
$(\pi,0)$, $(0,\pi)$ for the U band. The width of the peak is due
primarily to the overall strength of the z-axis couplings. This width
is very sensitive to the details of the structure. This sensitivity
controls the T dependence of the KS.
The other aspects of the density of states are robust.

Considering first the Cu spin relaxation, the two relevant orbitals
are $d_{x^2-y^2}$ and $d_{z^2}$. Because there is no orbital
relaxation between these orbitals regardless of the magnetic field
direction, the relevant relaxation is dipole-dipole. We will neglect
here the core polarization relaxation which we expect will not change
the qualitative conclusions. The relaxation rate for a z-axis 
magnetic field is given by,

$$^{63}W_z=2\biggl({2\pi\over\hbar}\biggr)(\gamma_e\gamma_h\hbar)^2
\int{\rm d}\ep f(\ep)(1-f(\ep))\biggl\langle{1\over r^3}\biggr\rangle^2
[W_{\rm dip}^z(\ep) + W_{\rm orb}^z(\ep)],\eqno(61)$$

$$W_{\rm dip}^z(\ep)=\biggl({1\over7^2}\biggr)
[6N_{d_{x^2-y^2}}(\ep)N_{d_{z^2}}(\ep) + 
N_{d_{x^2-y^2}}(\ep)N_{d_{x^2-y^2}}(\ep) +
N_{d_{z^2}}(\ep)N_{d_{z^2}}(\ep)
],\eqno(62)$$
$$W_{\rm orb}^z(\ep)=0,\eqno(63)$$

\noindent where $N(d_{x^2-y^2})(\ep)$ and $N(d_{z^2})(\ep)$ are the
total bare density of states at energy $\ep$ and $f(\ep)$ is the
Fermi-Dirac function.

The relaxation rate for a planar field $^{63}W_{xy}$ is
identical to $^{63}W_z$ above with $W_{\rm dip}^z$ and 
$W_{\rm orb}^z$ replaced with,

$$W^{xy}_{\rm dip}=
\biggl({1\over7^2}\biggr)\biggl({5\over2}\biggr)
[N_{d_{x^2-y^2}}(\ep)N_{d_{x^2-y^2}}(\ep) +
N_{d_{z^2}}(\ep)N_{d_{z^2}}(\ep)]\quad+$$
$$\biggl({3\over7^2}\biggr)
N_{d_{x^2-y^2}}(\ep)N_{d_{z^2}}(\ep),\eqno(64)$$
$$W_{\rm orb}^{xy}=0,\eqno(65)$$

Figures $4(a,b)$ show the amount of orbital character of $d_{x^2-y^2}$ and
$d_{z^2}$ in the U and L bands. The total bare density of states for
$d_{z^2}$ say, 
is the amount of orbital character of $d_{z^2}$ in the U band times
the density of states of the U band $N_U(\ep)$ plus a corresponding
term for the L band. Figures $6(a-c)$ 
are plots of the bare density of states
for each orbital due to each band and the total bare density of
states. In all calculations of the nmr and Hall effect, the
temperature dependence of chemical potential has been taken into
account. $\mu$ decreases about $5\times10^{-3}$ eV from $T=0$ to $T=300$K.
Finally, plots of $^{63}W_{z}/T$ and $^{63}W_{xy}/T$ are
shown in figures $7(a,b)$ where we have used the value$^{21}$
$\langle1/r^3\rangle=6$ a.u. The
curves show the characteristic peak at a T value greater than 
T$_c$ and the order
of magnitude of the rate is consistent with experiment.$^7$ With the above
figures and expressions for the relaxation rates we can qualitatively
see why the correct behavior is obtained. 

There are two different ways an electron may relax a Cu nucleus: one
by intra-band (U electron to U electron or L electron to L electron),
or two, by the inter-band processes U to L and L to U. Due to the
larger density of states of the U band, intra-band L to L relaxation
is small. This leaves the relaxation rate to be determined by the U
to U intra-band scattering and U to L (L to U) inter-band
scattering. The U to U scattering leads to an increase of $W/T$ due to
the sharp increase in the U band density at the saddle point
singularity. On the other hand, the contribution from the inter-band
term must decrease due to the closeness of the Fermi energy to the
very top of the L band and therefore the vanishing of the density of
states. The competition of these two terms gives the final result
shown. Regarding the Cu KS, only the U to U contribution and L to L
contribution can appear because the KS comes from diagonal elements
of the electron-nuclear Hamiltonian. The U to U KS contribution
dominates the L to L KS contribution due to the larger density of
states of the U band as before leading to a monotonically increasing
KS if the hyperfine couplings have the correct sign. The
sign of the couplings will be discussed later on, but we can now see
that the Cu KS and $W/T$ will have different temperature dependencies
as observed. 

It is pleasing to find that our calculated value for the relaxation
rate has the correct order of magnitude (experimentally,
$^{63}W_z/T\approx20$ s$^{-1}{\rm K}^{-1}$ at $100$K versus our value of
7.2 s$^{-1}{\rm K}^{-1}$).
Our calculation has completely neglected the effects of
the Cu 4s contact term and secondarily, the core polarization
contribution. Also, our calculated percentage increase from 30K to the
peak is about 3\% versus an experimental increase of $\approx 10-20\%$
and the percentage decrease from the peak value to the 300K value is
15\% versus the observed $\approx50\%$. In spite of these
differences, the qualitative behavior is correct and this is the most
important aspect given the level of calculation used in deriving the
Hubbard model parameters.

As we can see from figures $7(a,b)$, the relaxation anisotropy of $\approx3.4$
is not accounted for with the present model. What is missing here is a
small amount of $d_{xy}$ orbital character in our bands. $d_{xy}$ is
the next most unstable Cu orbital by ligand field theory after
$d_{x^2-y^2}$ and $d_{z^2}$. Including some $d_{xy}$ character affects
$W_{xy}$ dramatically because now orbital relaxation is permitted
$W^{xy}_{\rm orb}\ne0$, whereas $W^{z}_{\rm orb}$ remains zero. Including
$d_{xy}$ makes,

$$W^{xy}_{\rm orb}(\ep)=
4N_{d_{x^2-y^2}}(\ep)N_{d_{xy}}(\ep),\eqno(66)$$ 

$$W^z_{\rm orb}=0.\eqno(67)$$

\noindent There are also additional terms due to $d_{xy}$ in the
dipole-dipole terms (62), (64) but these are small and cannot account
for the anisotropy. We neglect them here.

The coefficient of 4 in front of $W^{xy}_{\rm orb}$ is about 30 times larger
than the first coefficient in (62) and more than two orders of
magnitude greater than the other coefficients
in $W^z_{\rm dip}$. Thus, $d_{xy}$ character on the order of a few
percent will lead to a contribution to $^{63}W_{xy}$ as large as
the dipole term leading easily to an anisotropy factor commensurate
with experiment.

For the planar O sites, the relevant orbital is $p_\sigma$. The O 2s
will be considered later.
The lack
of a second major O orbital near the Fermi surface and also the fact
that at $(\pi,\pi)$ there is no mixing of antibonding (most unstable)
$p_\sigma$ (symmetry
B$_{1g}$) with $d_{z^2}$ (A$_{1g}$) are the differences between O and
Cu that leads to the different spin relaxation rate temperature
dependencies. Figure $4c$ shows the amount of $p_\sigma$ character on an
O site at various energies. Figure $6c$ shows the $p_\sigma$ bare density
of states due to the U and L bands and the total bare density of
states. 

We can see that the contribution to the bare density of states due to
the L band $N_{L,p_\sigma}(\ep)$ is small compared to the U band term
$N_{U,p_\sigma}(\ep)$ leading to a small contribution to the
relaxation rate due to the L band. This is due to the L band k states
near the Fermi energy being close to $(\pi,\pi)$. At $(\pi,\pi)$,
$d_{z^2}$ couples to the neighboring $\sigma$ orbitals which must be
in a bonding configuration. This bonding set of
$\sigma$ orbitals is stabilized by the $t_{\sigma\sigma}$ term. 
Thus, to create
the most unstable $d_{z^2}$ configuration at $(\pi,\pi)$, we cannot
have a large amount of $p_\sigma$ character leading to the small value
for $N_{L,p_\sigma}(\ep)$ near the Fermi energy.

There are three distinct directions for the magnetic field at the O
site with different relaxation rates. These are the z-axis normal to
the CuO planes, the $\sigma$ axis along the Cu-O bond direction, and
the perpendicular axis $\perp$ normal to z and $\sigma$. With only the
$p_\sigma$ to consider, the relaxation along z and $\perp$ are equal,
$^{17}W_z=^{17}W_\perp\ne^{17}W_\sigma$. 

The expressions for $^{17}W_z$, $^{17}W_\sigma$ are of the same form
as for Cu (61) with the crude approximation
$\langle1/r^3\rangle\approx3$ a.u. determined by
ab-initio calculations and $\gamma_n$ the gyromagnetic ratio of the
$^{17}O$ nucleus. With only a $p_\sigma$ contributing, there is no
orbital relaxation in any direction of the magnetic field,

$$^{17}W^z_{\rm orb}=^{17}W^\sigma_{\rm orb}=
^{17}W^\perp_{\rm orb}=0.\eqno(68)$$

\noindent The dipole-dipole relaxation term is,

$$W^\sigma_{\rm dip}=\biggl({1\over5^2}\biggr)
N_{p_\sigma}(\ep)N_{p_\sigma}(\ep),\eqno(69)$$

$$W^z_{\rm dip}=W^\perp_{\rm dip}=
\biggl({5\over2}\biggr)W^\sigma_{\rm dip},\eqno(70)$$

\noindent A plot of the $^{17}W_\sigma$ relaxation is shown in figure
8. The
others are simply 2.5 times larger. This relaxation is monotonic
increasing due to the bare density of $p_\sigma$ due to the L band
being small. 

There remains one orbital that can make a large contribution
to the relaxation due to its large hyperfine coupling to the nucleus
that we have not yet considered. That is the O 2s
orbital. A very small amount of 2s can have a large effect on $^{17}W$
without making any change in our calculated band structure. At
$(\pi,\pi)$, the L band has no $d_{x^2-y^2}$ character and by symmetry
no 2s character either. Thus, similar to $N_{Lp_\sigma}(\ep)$, there
is almost no contribution to the O 2s bare density $N_{2s}(\ep)$ due
to the L band leading to the same monotonic increasing behavior as in
figure 8. O 2s will make a large change to the absolute value of the
relaxation rate. Without estimating the size of the 2s contribution,
we cannot compare our answer in figure 8 to experimental data on
LASCO. Without O 2s, the values in figure 8 are about an order of
magnitude too small.$^7$

The KS on the O sites is due to the O 2s Fermi contact interaction,
the $p_\sigma$ orbital dipole coupling and core polarization. In
figure 9, we simply plot the KS,

$$K_\sigma=-2K_z=-2K_\perp=
\biggl({8\over5}\biggr)\biggl\langle{1\over r^3}\biggr\rangle
\mu_B^2\int\biggl(
-{\partial f\over\partial\ep}\biggr)
N_{p_\sigma}(\ep)d\ep,\eqno(71)$$

\noindent due to the $p_\sigma$ dipole-dipole term. Since the 2s and
core polarization shifts are isotropic, our curve may be compared to
the axial KS, $K_{\rm ax}=2(K_\sigma-K_\perp)/3$.
The monotonic increase in the shift is due to the sharp increase in
the U band density of states above $\ep_F$ due to the saddle point
singularity at $(\pi,0)$, $(0,\pi)$.

The KS involves only one factor of $N_{p_\sigma}(\ep)$ while the
relaxation contains $N_{p_\sigma}(\ep)^2$. Thus, there is no way
$^{17}W/T \propto K_\sigma$ is strictly possible, although from the figures,
the agreement is not far off.

For the Cu KS, we need to introduce the effects of the Cu 4s orbital
and the two $p_z$ orbitals above and below the Cu on the apical
oxygens. Experimentally, the Cu KS for a z-axis field is T independent
while for planar fields the shift is monotonic increasing with
increasing T. As we discussed, the dominant contribution to the shift
will come from the bare density of states for the orbitals from the U
bands. If we can somehow show that the hyperfine couplings are
positive for planar fields and zero for the z-axis field, then we
should expect a curve similar to figure 9 for Cu KS.

The Cu KS along the z-axis and planar directions is the sum of,

$$^{63}K_\alpha=K_\alpha^{\rm dip}(d_{x^2-y^2}) +
K_\alpha^{\rm dip}(d_{z^2}) +
K_\alpha^{\rm contact}(4s) +
K_\alpha^{\rm cp} +
K_\alpha^{\rm dip}(d_{z^2},4s),\eqno(72)$$

\noindent where $\alpha$ is the field direction z or planar, $K^{\rm
dip}(d_{x^2-y^2})$, $K^{\rm dip}(d_{z^2})$ are the dipole shifts due
to the $d_{x^2-y^2}$ and $d_{z^2}$ orbitals, $K^{\rm contact}(4s)$ is
the Fermi contact shift, $K^{\rm cp}$ is the core polarization shift,
and $K^{\rm dip}(d_{z^2},4s)$ is the dipole contribution due to the
interference of $d_{z^2}$ and 4s. The expressions for the first four
terms are,

$$K^{\rm dip}_z(d_{x^2-y^2})=-2K^{\rm dip}_{xy}(d_{x^2-y^2})=
-{8\over7}\biggl\langle{1\over r^3}\biggr\rangle
\mu_B^2\int\biggl(-{\partial f\over\partial\ep}\biggr)
N_{d_{x^2-y^2}}(\ep)d\ep,\eqno(73)$$

$$K^{\rm dip}_z(d_{z^2})=-2K^{\rm dip}_{xy}(d_{z^2})=
+{8\over7}\biggl\langle{1\over r^3}\biggr\rangle
\mu_B^2\int\biggl(-{\partial f\over\partial\ep}\biggr)
N_{d_{z^2}}(\ep)d\ep,\eqno(74)$$

$$K^{\rm contact}_z(4s)=K^{\rm contact}_{xy}(4s)=
{16\pi\over3}|\psi_{4s}(0)|^2
\mu_B^2\int\biggl(-{\partial f\over\partial\ep}\biggr)
N_{4s}(\ep)d\ep,\eqno(75)$$

$$K^{\rm cp}_z=K^{\rm cp}_{xy}=
-(2\alpha)\biggl\langle{1\over r^3}\biggr\rangle
\mu_B^2\int\biggl(-{\partial f\over\partial\ep}\biggr)
[N_{d_{x^2-y^2}}(\ep)+N_{d_{z^2}}(\ep)]d\ep.\eqno(76)$$

\noindent We take the value $\alpha=0.33$ (dimensionless) from Abragam
and Bleaney.$^{21}$ $\psi_{4s}(0)$ is the value of the 4s orbital at the
nucleus.

The $d_{z^2}$, 4s interference term is evaluated by taking the mean
value of a band wavefunction at the Fermi surface with the dipole
Hamiltonian. Let,

$$\phi_k=A_kd_{x^2-y^2k} +B_kd_{z^2k} + C_k\psi_{4sk} +
\ldots,\eqno(77)$$ 

\noindent where $d_{x^2-y^2k}$, $d_{z^2k}$ are defined in (34), (35) and
$\psi_{4sk}$ is defined similarly. Then,
$$\leftline{\indent$\displaystyle
\langle\phi_k|H_{\rm dip}|\phi_k\rangle=
(-2\mu_B)(\gamma_n\hbar)\bigl[
I_xS_x+I_yS_y-2I_zS_z\bigr]\cdot$}$$
$$\biggl\{
\biggl(-{2\over7}\biggl\langle
{1\over r^3}\biggr\rangle\biggr)|A_k|^2+
\biggl(+{2\over7}\biggl\langle{1\over r^3}\biggr\rangle\biggr)
|B_k|^2 +
\biggl({1\over\sqrt5}
\biggl\langle{1\over r^3}\biggr\rangle_{z^2,s}\biggr)
\bigl(B_k^*C_k+B_kC_k^*\bigr)\biggr\}$$
$$+\ \hbox{term involving}\  (A_k^*C_k+A_kC_k^*),\eqno(78)$$

$$\biggl\langle{1\over r^3}\biggr\rangle_{z^2,s}=
\int_0^{+\infty}r^2drR_{d_{z^2}}(r)\biggl({1\over r^3}\biggr)
R_{4s}(r),\eqno(79)$$

\noindent where $I$ is the nuclear spin and
$R_{d_{z^2}}(r)$, $R_{4s}(r)$ are the radial parts of
the $d_{z^2}$ and 4s orbitals respectively with normalizations,

$$\int_0^{+\infty}r^2drR_{d_{z^2}}(r)^2=
\int_0^{+\infty}r^2drR_{4s}(r)^2=1.\eqno(80)$$

Averaging over the Fermi surface, the interference term $A_kC_k^*+
A_k^*C_k$ due to $d_{x^2-y^2}$ and 4s becomes zero due to the
different symmetries of $d_{x^2-y^2}$ and 4s (B$_{1g}$ and
A$_{1g}$). The first two terms in (78) lead to
$K^{\rm dip}(d_{x^2-y^2})+K^{\rm dip}(d_{z^2})$ and the third term
gives, 

$$K^{\rm dip}_z(d_{z^2},4s)=
\biggl({8\over\sqrt5}\biggr)
\biggl\langle{1\over r^3}\biggr\rangle_{z^2,s}
\mu_B^2\int\biggl(-{\partial f\over\partial\ep}\biggr)
\bigl\langle B_k^*C_k\bigr\rangle N(\ep){\rm d}\ep,\eqno(81)$$
$$K^{\rm dip}_{xy}(d_{z^2},4s)=-\o2
K^{\rm dip}_z(d_{z^2},4s),\eqno(82)$$

\noindent where $<B_k^*C_k>$ is the mean value of $B_k^*C_k$ over the
Fermi surface and $N(\ep)$ is the total density of states of the
band. Depending on the sign of $<B_k^*C_k>$, the shift due to the
interference term can be positive enhancing the field due to
$d_{z^2}$, or negative decreasing the net magnetic field of
$d_{z^2}$. The bare $d_{z^2}$ and $4s$ density of states from this
band are $\langle|B_k|^2\rangle N(\ep)$ and $\langle|C_k|^2\rangle
N(\ep)$. Since only one power of $C_k$ appears in (81), 
$\langle B_k^*C_k\rangle$ can be large although
$\langle|C_k|^2\rangle$ may be small.

The bare $d_{x^2-y^2}$ density of states is smaller than the $d_{z^2}$
density of states at the Fermi energy leading to a net positive shift
from $K(d_{x^2-y^2})+K(d_{z^2})$ along the z-direction and a net
negative shift along the plane. $K^{\rm contact}(4s)$ is isotropic and
always positive.

At all k vectors on the Fermi surface, the $d_{z^2}$ and $p_z$'s on
the apical O form an antibonding combination leading to a larger
amount of 4s character than one expects from a $d_{x^2-y^2}$
band. Thus, if the $d_{z^2}$, 4s interference term leads to a negative
$<B_k^*C_k>$ that is sufficiently large, the net effect of $K(d_{x^2-y^2})+
K(d_{z^2}) + K(d_{z^2},4s)$ leads to a dipolar field which is negative
in the z direction and positive along the plane. We shall assume that
due to the charge donation of the apical O $p_z$ to the Cu 4s, this is
the case.
With this assumption, we can see how the net shift on the Cu due to a
z-axis field can add to zero while simultaneously leading to positive
shifts for planar fields. In (72), the contribution to $K_\alpha$ from
the contact and core polarization terms should be a net isotropic
positive value. Without the interference shift (81), there is no way
the z-axis shift can add to zero.
Figure 10 shows the T dependence of the z-axis shift of
$K(d_{x^2-y^2})$ and $-K(d_{z^2})$. These curves have approximately
the same T dependence as the O shift and $^{17}W/T$. 

The above considerations also show why for fully doped YBCO$_7$,
the T dependence of the shifts and $^{17}W/T$ are almost constant. If
there is more dispersion in the z-direction of the $d_{z^2}$ orbital,
then the saddle point peak in the U band density of states will be
broadened. If the plateau is sufficiently broad to extend all the way
to the Fermi energy or the $(\pi,0)$ and $(0,\pi)$ saddle point
singularity is sufficiently far away from the Fermi energy,
then the shifts and O relaxation will become T
independent. For optimally doped YBCO$_7$, there are no vacancies in
the chains and one can expect more 3D dispersion from the $d_{z^2}$
orbital. 
\medskip
\noindent{\bf B. Hall Effect and Resistivity}

Since the density of states for the L band is smaller than for the U
band in the vicinity of the Fermi energy at the crossing point, we
take the L band as the primary carrier of current. The Hall
coefficient $R_H$ is the ratio of the transverse conductivity
$\sigma_{xy}$ and the conductivity $\sigma$ squared. Using standard
Boltzman theory,$^{22}$ 

$$\sigma_{xy}=\biggl({m_e^2\Omega_0\over\hbar}\biggr)
\biggl({1\over\Omega}\biggr)\sum_k
\biggl(-{\partial f\over\partial\ep_k}\biggr)
v_{k_y}\biggr[v_{k_y}{\partial\over\partial k_x}-
v_{k_x}{\partial\over\partial k_y}\biggr]v_{k_x},\eqno(83)$$

$$\sigma=m_e\Omega_0\biggl({1\over\Omega}\biggr)\sum_k
\biggl(-{\partial f\over\partial\ep_k}\biggr)
v_{k_x}^2,\eqno(84)$$

$$R_H=\biggl({\Omega_0\over qc}\biggr)
{\sigma_{xy}\over\sigma^2},\eqno(85)$$

\noindent where $m_e$ is the electron mass, $\Omega_0$ is the
primitive unit cell volume $\Omega_0=96$ \AA, $\Omega$ is the total
volume, $q=-|e|<0$ is the electron charge, and $c$ is the velocity of
light. $v_k=\nabla_k\ep_k/\hbar$ is the velocity.
We have multiplied $\sigma_{xy}$ and $\sigma$ by the
appropriate factors of $m_e$ and $\Omega_0$ to make the expressions in
(82) and (83) dimensionless and have neglected the
scattering rate $1/\tau$ in these expressions because for $R_H$,
$\tau$ does not appear.
We may also define $\sigma_{xy}(\ep)$ and
$\sigma(\ep)$ by replacing the Fermi-Dirac function 
$-\partial f/\partial\ep_k$ by the delta function $\delta(\ep_k-\ep)$
in (82) and (83). Then,

$$\sigma_{xy}=\int\sigma_{xy}(\ep)
\biggl(-{\partial f\over\partial\ep_k}\biggr)
d\ep,\eqno(86)$$

$$\sigma=\int\sigma(\ep)
\biggl(-{\partial f\over\partial\ep_k}\biggr)
d\ep.\eqno(87)$$

Figure 11 is a plot of $\sigma_{xy}(\ep)$ and $\sigma(\ep)$ and figure
12 shows the temperature dependence of the Hall coefficient at optimal
doping. The absolute magnitude of $R_H$ is about ten times larger than
observed values for LASCO, but the calculated percentage change of
$R_H$ from the 100K to 300K is $\approx40$\% 
in good agreement with experiment$^{12}$ ($\approx50$\%).  

These
curves were calculated taking the small z-axis dispersion parameters
$T_{(0,0)}$, $T_{(\pi,0)}$, and $T_{(\pi,\pi)}$ in $(31)-(33)$ to be
zero. This was done to simplify the computation.
$\sigma_{xy}(\ep)$ changes very rapidly for energies higher than
the Fermi energy at the band touching point, whereas the change in
$\sigma(\ep)$ is not as dramatic. 

The abrupt change in $\sigma_{xy}(\ep)$ is
due to the two band crossing point. Near the Fermi energy of the
crossing point, the repulsion of the two bands for non-diagonal $k$
vectors strongly affects the shape of the two Fermi surfaces leading to
a large curvature for the L band. At energies slightly higher than the
Fermi energy, 
the L band surface is not affected by the presence of the U band and
the Fermi surface becomes ``small'' leading to the reduced
curvature. The overall T dependence of $R_H$ is due to the combined T
dependencies of both $\sigma$ and $\sigma_{xy}$ rather than solely due
to $\sigma_{xy}$ as one would expect from a first glance at figure 11.
We can see that the presence of
the two-band crossing point and the association of optimal
superconductivity with this point is the root cause of the strong
anomalous temperature dependence of the Hall coefficient.

There are several possible explanations for our calculated value for
$R_H$ being more than ten times too large. The most obvious one is
the neglect of next-nearest neighbor hopping terms in our Hubbard
model. The second cause is due to our 2D approximation to the band
structure. 
A more detailed description than ours of the z-axis
dispersion is required. The third reason is the neglect of the
contribution from the electron-like surface of the U band.
Finally,
more refined ab-initio calculations than the ones performed in the
following paper on the CuO$_6$ complex embedded in the point charge
field of LASCO will change the parameters used in our Hubbard model.

The resistivity due to scattering with phonons 
should be linear at the optimal doping for two
reasons: 1.) the Fermi surface for the L band is ``small'', and 2.) at
optimal doping the electron current can strongly relax by scattering
to a U electron state. At optimal doping, the U band Fermi surface
touches the L band surface leading to nearby states in k space with
very different currents. The T dependence of $\sigma$ complicates this
scenario because $\sigma$ is proportional to the effective number of
charge carriers in the band. It is unclear
how the T dependence of the relaxation rate
$1/\tau$ could cancel this dependence. 
If a full-blown 3D model was used instead of our 2D model with the
third dimension included as a perturbation, we believe that the
temperature dependence of $\sigma$ would become small while
$\sigma_{xy}$ would remain very T dependent.
\bigskip
\centerline{\bf CONCLUSIONS}

We have presented a model for superconductivity of the cuprates based
on the idea that Cooper pairs are formed from electrons between two
distinct bands. This leads naturally to associating optimal doping
with a Fermi surface touching point of the two bands. We have
postulated the character of the two bands to arise from $d_{x^2-y^2}$,
$d_{z^2}$, O $p_\sigma$, and apical O $p_z$ orbitals. A Hubbard model
for these bands is set-up and we calculate some normal state
consequences of the model. 
With our model, many of the anomalous features of the normal state nmr,
Hall effect, and resistivity are explained qualitatively and to
varying degrees quantitatively. The primary reason for the anomalous
normal state properties is due to the optimal doping being at the
Fermi surface touching point.

We show that inter-band pairing alters the standard interpretation of
Josephson tunneling. With inter-band pairs, the detailed nature of the
single particle tunneling matrix elements plays a prominent role. We
show that with this new piece of physics, three of the four Josephson
tunneling experiments are explained by our model with a phonon
mediated attractive coupling.

The parameters in the Hubbard model used in this paper
are derived in the following
paper by Perry and Tahir-Kheli
from calculations on a CuO$_6$ cluster for
La$_{2-x}$Sr$_x$CuO$_4$. We conclude there that contrary to band
structure calculations where only one band with Cu $d_{x^2-y^2}$ and O
$p_\sigma$ character is found, two bands exist at the Fermi energy
with the character described above.
\bigskip\bigskip
\vfil\eject
\noindent{\bf REFERENCES}\smallskip
\item {1.} J. Tahir-Kheli, in {\it Proceedings of the 10th Anniversary
HTS Workshop on Physics, Materials and Applications}, edited by
B. Batlogg, C.W. Chu, W.K. Chu, D.U. Gubser, and K.A. M\"uller,
(World Scientific, Singapore 1996), p. 491
\item {2.} D.A. Wollman, D.J. van Harlingen, W.C. Lee, D.M. Ginsberg,
and A.J. Leggett, {\it Phys. Rev. Lett.}
{\bf 71}, 2134 (1993)
\item {3.} C.C. Tsuei, J.R. Kirtley, C.C. Chi, Lock See Yu-Jahnes,
A. Gupta, T. Shaw, J.Z. Sun, and M.B. Ketchen,
{\it Phys. Rev. Lett.} 
{\bf 73}, 593 (1994)
\item {4.} J.K. Perry and J. Tahir-Kheli,
http://www.firstprinciples.com, cond-mat/9711184
\item {5.} M. Takigawa, P.C. Hammel, R.H. Heffner, Z. Fisk, K.C. Ott,
and J.D. Thompson,
{\it Phys. Rev. Lett.} {\bf 63}, 1865
(1989) 
\item {6.} M. Takigawa, P.C. Hammel, R.H. Heffner, Z. Fisk, J.L. Smith,
and R.B. Schwarz,
{\it Phys. Rev. B} {\bf 39}, 300 (1989)
\item {7.} R.E. Walstedt, B.S. Shastry, and S-W. Cheong,
{\it Phys. Rev. Lett.} {\bf 72}, 3610
(1994)
\item {8.} T. Imai, K. Yoshimura, T. Uemura, H. Yasuoka, K. Kosuge,
{\it J. Phys. Soc. Jpn.} {\bf 59}, 3846
(1990)
\item {9.} M. Takigawa, A.P. Reyes, P.C. Hammel, J.D. Thompson,
R.H. Heffner, Z. Fisk, and K.C. Ott,
{\it Phys. Rev. B} {\bf 43}, 247 (1991)
\item {10.} R.E. Walstedt, W.W. Warren, Jr., R.F. Bell, G.F. Brennert,
G.P. Espinoza, R.J. Cava, L.F. Schneemeyer, and J.V. Waszczak,
{\it Phys. Rev. B} {\bf 38}, 9299
(1988) 
\item {11.} M. Horvati\'c, C. Berthier, P. S\'egransan, P. Butaud,
W.G. Clark, and J.A. Gillet,
{\it Phys. Rev. B} {\bf 48}, 13848
(1993)
\item {12.} H.Y. Hwang, B. Batlogg, H. Takagi, H.L. Kao, J. Kwo,
R.J. Cava, J.J. Krajewski, and W.F. Peck, Jr.,
{\it Phys. Rev. Lett.} {\bf 72}, 2636
(1994) 
\item {13.} A.G. Sun, D.A. Gajewksi, M.B. Maple, and R.C. Dynes,
{\it Phys. Rev. Lett.} {\bf 72}, 2267
(1994) 
\item {14.} P. Chaudhari and Shawn-Yu Lin,
{\it Phys. Rev. Lett.} {\bf 72}, 1084
(1994)
\item {15.} H. Takagi, R.J. Cava, M. Marezio, B. Batlogg,
J.J. Krajewski, and W.F. Peck, Jr.,
{\it Phys. Rev. Lett.} {\bf 68},
3777 (1992).
\item {16.} R.J. Cava, B. Batlogg, C.H. Chen, E.A. Rietman,
S.M. Zahurak, and D. Werder,
{\it Nature} {\bf 329}, 423 (1987).
\item {17.} H. Takagi, B. Batlogg, H.L. Kao, J. Kwo, R.J. Cava,
J.J. Krajewski, and W.F. Peck, Jr.,
{\it Phys. Rev. Lett} {\bf 69}, 2975
(1992)
\item {18.} P.G. deGennes, {\it Superconductivity of Metals and
Alloys}, (Addison-Wesley, 1966)
\item {19.} Z.-X. Shen, D.S. Dessau, B.O. Wells, D.M. King,
W.E. Spicer, A.J. Arko, D. Marshall, L.W. Lombardo, A. Kapitulnik,
P. Dickinson, S. Doniach, J. DiCarlo, A.G. Loeser, and C.H. Park,
{\it Phys. Rev. Lett.} {\bf 70}, 1553
(1993) 
\item {20.} H. Ding, A.F. Bellman, J.C. Campuzano, M. Randeria,
M.R. Norman, T. Yokoya, T. Takahashi, H. Katayama-Yoshida, T. Mochiku,
K. Kadowaki, G. Jennings, and G.P. Brivio,
{\it Phys. Rev. Lett.} {\bf 76}, 1533
(1996) 
\item {21.} A. Abragam and B. Bleaney, {\it Electron Paramagnetic
Resonance of Transition Ions} (Dover, 1986), p. 458
\item {22.} J.M. Ziman, {\it Electrons and Phonons}, (Oxford
University Press, 1960)

\vskip0.5truein
\noindent {\bf Table 1.} Parameters in Hubbard Model in (eV).
\medskip
$$\vbox{\offinterlineskip
\hrule\vskip1pt
\hrule
\halign
   {\strut#&\quad #\hfil\quad &\hfil #\quad&#\cr
        &$\ep_{d_{x^2-y^2}}$&$-2.403$&\cr
        &$\ep_{d_{z^2}}$&$-2.092$&\cr
        &$\ep_{p_\sigma}$&$ -6.122$&\cr
        &$\ep_{p_z}$&$-0.852$&\cr
        &$t_{x^2-y^2,\sigma}$&$1.347$&\cr
        &$t_{z^2,\sigma}$&$0.514$&\cr
        &$t_{\sigma\sigma}$&$0.368$&\cr
        &$t_{\sigma\sigma a}$&$-0.041$&\cr
        &$t_{p_z,d_{z^2}}$&$1.076$&\cr
        &$t_{p_z,\sigma}$&$0.078$&\cr
        &$t_{p_z,p_z}$&$0.493$&\cr
}        
\vskip2pt\hrule\vskip1pt\hrule}$$
\vfil\eject
\end